\documentclass[aps,pre,twocolumn,floatfix,showpacs,groupaddress,longbibliography,nofootinbib]{revtex4-2}

\usepackage{amssymb}
\usepackage{amsmath}
\usepackage{bm}
\usepackage{graphicx}
\usepackage{hyperref}
\usepackage{xcolor}
\usepackage{soul}

\DeclareMathOperator{\Arctan}{Arctan}

\DeclareMathOperator{\im}{Im}
\DeclareMathOperator{\re}{Re}
\usepackage[utf8]{inputenc}
\newcommand{\zb}{\bar{z}}

\begin{document}

\title{Defect Interactions Through Periodic Boundaries in Two-Dimensional $p$-atics}

\author{Cody D. Schimming}
\email[]{cschimm2@jh.edu}
\affiliation{Department of Physics and Astronomy, Johns Hopkins University, Baltimore, Maryland, 21211, USA}

\begin{abstract}
Periodic boundary conditions are a common theoretical and computational tool used to emulate effectively infinite domains.
However, two-dimensional periodic domains are topologically distinct from the infinite plane, eliciting the question: How do periodic boundaries affect systems with topological properties themselves?
In this work, I derive an analytical expression for the orientation fields of two-dimensional $p$-atic liquid crystals, systems with $p$-fold rotational symmetry, with topological defects in a flat domain subject to periodic boundary conditions. 
I show that this orientation field leads to an anomalous interaction between defects that deviates from the usual Coulomb interaction, which is confirmed through continuum simulations of nematic liquid crystals ($p = 2$). 
The interaction is understood as being mediated by non-singular topological solitons in the director field which are stabilized by the periodic boundary conditions.
The results show the importance of considering domain topology, not only geometry, when analyzing interactions between topological defects.
\end{abstract}

\maketitle

\section{Introduction} \label{sec:Intro}

Topological defects are topologically protected excitations that arise in systems with spontaneous symmetry breaking \cite{Mermin79,chaikin95}. 
Examples of topological defects manifest in biological, material, high-energy, and cosmological physics and include monopoles, disclinations, dislocations, vortices, Skyrmions, Hopfions, and cosmic strings \cite{abrikosov57,halperin78,Defects94,blatter94,kibble97,pismen99,kleman08,Ackerman17,duclos20,Maroudas-Sacks21,Tai24}. 
In biological physics, defects may execute biological functions such as cell apoptosis and morphogenesis \cite{livshits17,saw17,Maroudas-Sacks21,hoffman22}; in material physics, defects may be manipulated to alter a material's optical and mechanical properties, facilitate self-assembly, actuate surfaces, or perform computations \cite{peach50,escaig74,ravnik07,Wang2016-yc,baba18,guo21,RZhang22}; while in cosmological physics, defects are predicted to interact with light and matter, but have yet to be observed \cite{kibble97,Servant24}. 
Thus, understanding the dynamics of defects is of fundamental importance to many areas of physics.

One of the simplest systems that admits topological defects are the $p$-atic liquid crystals, which describe systems with $p$-fold rotational symmetry.
In two dimensions, the most common $p$-atics are those with $p = 1,2,6$, corresponding to the continuous $XY$-model, nematic liquid crystals, and hexatic liquid crystals.
While these three models have been heavily studied \cite{chaikin95}, there have been recent experimental realizations of other $p$-atics \cite{Zhao2012-ik,Wang2018-yu,Meijer2019-ye}, as well as recent theoretical work involving the hydrodynamics, anyonic braiding, and interaction with curvature of arbitrary $p$-atic defects \cite{Giomi22Lett,Giomi22,Mietke22,Vafa22b}.

The dynamics of $p$-atics, both in equilibrium and non-equilibrium settings (and particularly when topological defects are involved), are often studied using computational models.
A common choice for boundary conditions in computations is periodic boundary conditions (PBC), which are often employed to emulate an effectively infinite system.
In the infinite plane, it is known that topological defects interact via Coulomb forces, akin to electrostatic charges, for systems with a single elastic constant \cite{chaikin95}.
The Coulomb interaction between defects has also been observed experimentally in passive nematic liquid crystals \cite{zushi22}.
However, since domains with PBC are topologically distinct from infinite systems, it is not obvious that the Coulomb interactions in the infinite plane should be the same as interactions in PBC.
There have been several prior analytical and computational studies of topological defects on the torus, a domain topologically identical to PBC, which have found anomalous interactions between defects \cite{Bowick04,Selinger2011-gm,ellis18,Rojo-Gonzalez24}, though this is typically attributed to curvature, and to my knowledge the role of topology (as opposed to curvature) in these interactions has not yet been clearly ascertained.

Here, I study $p$-atic topological defect interactions in flat systems with PBC.
In Sec. \ref{sec:Defects} I lay the theoretical foundation regarding defect configurations in two dimensional $p$-atics and review the relevant results regarding defects in the infinite plane.
Then, in Sec. \ref{sec:PBC}, I attempt to derive the $p$-atic configuration with defects subject to PBC.
I show that if one assumes that defects behave like electrostatic particles (as they do in the infinite plane) the resulting configuration will not be periodic.
Instead, to conform to the boundary conditions, the configuration must be modified, which alters the interaction between defects to be highly asymmetric.
In Sec. \ref{sec:Solitons}, I show how this asymmetric interaction may be interpreted as being mediated by non-singular topological solitons that are stabilized by the PBC and how these results impact the analysis of computational studies.
Finally, in Sec. \ref{sec:Annulus} I conclude by positing how these asymmetric effects may play a role in physically realizable domains.

\section{$p$-atic topological defects in the plane} \label{sec:Defects}

$p$-atic liquid crystals are characterized by local orientational order with $p$-fold rotational symmetry.
In two dimensions, the orientational configuration of a $p$-atic may be summarized with an angle field $\theta_p(x,y)$, which represents the angle of the orientation at each point in space and it is understood that $\theta_p$ is defined modulo $2\pi/p$.
Variations in $\theta_p$ are penalized through an elastic free energy:
\begin{equation} \label{eqn:ElasticEnergy}
    F_{el} = \int \frac{K}{2}|\nabla\theta_p|^2\,d\mathbf{r}
\end{equation}
where $K$ is the elastic constant.
For $p=1$ or $p=2$ there may be two elastic constants, but I will take the one-constant approximation here for analytical tractability.
Thus, energy minimizing configurations must satisfy $\nabla^2\theta_p=0$.

Topological defects in two-dimensional $p$-atics are points where the orientation is singular.
They carry a winding number, or topological charge, defined as
\begin{equation} \label{eqn:DefectCharge}
    \frac{n}{p} = \frac{1}{2\pi}\oint_C\nabla\theta_p\cdot d\bm{\ell}
\end{equation}
where $n$ is an integer and $C$ is a closed loop.
Configurations that contain defects will thus be energy stabilizing if they satisfy $\nabla^2\theta_p = 0$ with Eq.~\eqref{eqn:DefectCharge} as a constraint.
For the rest of the paper I will assume, for simplicity, that only $\pm 1/p$ charge defects populate the system.
Higher charge defects are energetically unstable, unless stabilized by confinement or external fields.

It is often useful to describe the configuration using a complex-valued, analytic function $\Omega(z) = \phi(z,\bar{z}) - i p \theta_p(z,\bar{z})$ where $z = x + iy$ and $\phi$ is the harmonic conjugate of $\theta_p$ \cite{Chandler23,Chandler24,Copar24}.
Then $\nabla^2\theta_p = 0$ is automatically satisfied and topological defects are represented as poles of $\Omega$.
It will also be useful to identify $\phi(z,\bar{z}) = \re\left[\Omega(z)\right]$ as the electrostatic potential of an equivalent configuration in which the poles of $\Omega$ are electrostatic charges.

A single defect in the plane has the configuration
\begin{equation}
    \theta_p(z,\bar z) = k\im\left[\log(z-z_0)\right] + \theta_0
\end{equation}
where $k$ is the defect charge, $z_0 = x_0 + iy_0$ is the location of the defect, and $\theta_0$ is a global phase.
For a system with many defects, the configuration is given by the sum of single defect configurations, assuming all relative defect orientations minimize the free energy \cite{vromans16,tang17,Copar24}.
To analyze the interaction between defects I will follow the work of Halperin and Mazenko \cite{halperin81,liu92,mazenko97,mazenko99}, later adapted for liquid crystals \cite{angheluta21,schimming23}, which gives the defect velocity as a function of derivatives of the order parameter.
If defect motion is induced only by elastic relaxation (i.e. there are no flows or external fields) the velocity, $v = v_x +iv_y$, of the $j$th defect is \cite{angheluta21}
\begin{equation} \label{eqn:DefectVelocity}
    v_j = -\frac{4p^2k_jK}{\gamma}\left.i\partial_{\bar{z}}\tilde{\theta}_p\right|_{z=z_j}
\end{equation}
where $\partial_z \equiv (1/2)(\partial_x - i\partial_y)$, $\tilde{\theta}_p$ is the non-singular part of the configuration at $z_j$, $k_j$ is the defect charge, $K$ is the elastic constant, and $\gamma$ is the rotational viscosity (for the remainder of the paper I will work in units so that $K = \gamma = 1$).
If the multi-defect orientation 
\begin{equation}
\tilde{\theta}_p = \sum_{i\neq j}k_i\im\left[\log(z-z_i)\right] +\theta_0 
\end{equation}
is inserted into the expression, the result is a Coulomb interaction between defects:
\begin{equation}
    \bar{v}_j = 4p^2 \sum_{i\neq j} \frac{k_i k_j}{z_i - z_j}
\end{equation}
where the complex conjugate, $\bar{v}$, is used to directly compare with the electrostatic interaction between point charges (the complex electric field is often defined by $E = E_x - iE_y$) \cite{Hill21}.
This well-established result can also be derived using other methods \cite{deGennes75,chaikin95,pismen99}.
The result also follows conceptually from applying the Cauchy-Riemann equations to $\Omega(z)$, namely $i p \partial_z\theta_p = -\partial_z \phi$.
As discussed above, identifying $\phi$ as the electrostatic potential of the equivalent charge configuration reinterprets this as $\bar{v}\sim E$.
A natural question to ask is: can we treat $p$-atic defects as Coulomb charges in PBC? 

\begin{figure}
    \centering
    \includegraphics[width = \columnwidth]{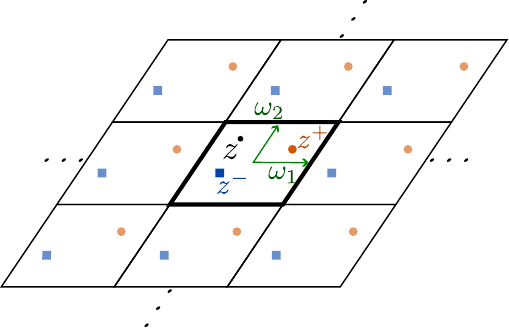}
    \caption{Schematic of the periodic system. Positive defects are located at $z^+$, negative defects at $z^-$, and a test defect at $z$. 
    $\omega_1$ and $\omega_2$ give the principal directions of the periodic lattice.}
    \label{fig:SystemSetup}
\end{figure}

\section{Periodic boundary conditions} \label{sec:PBC}

To handle PBC, the goal is to find a configuration $\theta_p$ that has the property $\theta_p(z + 2\ell\omega_1 +2m\omega_2) = \theta_p(z) \mod 2\pi/p$ where $\ell$ and $m$ are integers and $\omega_1$ and $\omega_2$ are complex numbers such that $\im \left[\omega_2/\omega_1\right] > 0$ which describe the shape and size of the periodic domain.
Without loss of generality, I will take $\omega_1$ to be real.
Note that, in general, the domain is a parallelogram [see Fig.~\ref{fig:SystemSetup}], but if $\omega_2/\omega_1$ is purely imaginary the domain is rectangular.
There are two immediate consequences of the periodicity: First, given a configuration, we may modify it to have any integer number of $2\pi/p$ rotations, $n_1,n_2$, along the $\omega_1,\omega_2$ cycles.
That is, given a periodic $\theta_p$,
\begin{equation} \label{eqn:GlobalSolitonAddition}
    \theta_p \to \theta_p + \frac{1}{p}\im\left\{\frac{2\pi i}{\omega_2 - \bar{\omega}_2}\left[-\frac{\omega_2}{\omega_1}n_1 + n_2\right]z\right\}
\end{equation}
will also be a periodic configuration.
The second consequence of the PBC is that the sum of all defect charges must be zero by the Poincar\'{e}-Hopf theorem.
Then, the defects populate the domain in pairs and a configuration may be constructed by taking a periodic tiling of the center cell, shown schematically in Fig.~\ref{fig:SystemSetup}.

\begin{figure}
    \centering
    \includegraphics[width = \columnwidth]{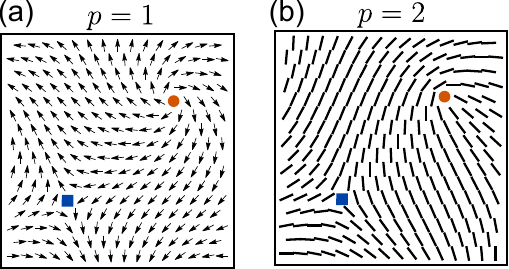}
    \caption{Orientation fields compatible with defects interacting as Coulomb charges, Eq.~\eqref{eqn:CoulombOrderParam}, for a system with $\omega_1 = 0.5$, $\omega_2 = 0.5i$, $z^+ = 0.25+0.25i$, $z^- = -0.25 - 0.25i$ and (a) $p=1$ or (b) $p=2$.
    In this case, periodic boundary conditions are not satisfied.}
    \label{fig:NonPeriodicConfiguration}
\end{figure}

\subsection{Failure of the electrostatic analogy}

To find a periodic configuration, I will first attempt to make an analogy with electrostatics, since this analogy holds in the case of the infinite plane.
To determine the configuration, I place a ``test defect'' at location $z$, which is meant to measure the force induced by the defect configuration already present in the domain, $\theta_p(z,\bar{z})$.
Following the expectation that $p$-atic defects interact like Coulomb charges, I assume that the force on the test defect is equivalent to the electric field of the corresponding point charge configuration, that is, I will assume $\bar{v}(z) = E(z)$ \cite{Hill21}.
Additionally, a useful identity is 
\begin{equation} \label{eqn:OmegaIdentity}
    \partial_z\Omega = -2i p\partial_z\theta_p
\end{equation}
which follows from the Cauchy-Riemann equations.
Using Eqs.~\eqref{eqn:DefectVelocity} and \eqref{eqn:OmegaIdentity}, $\Omega(z)$ satisfies
\begin{equation} \label{eqn:CoulombVelocity}
\partial_z\Omega = -\sum_{j=1}^N\left[\zeta(z-z_j^+) - \zeta(z-z_j^-) + f(\Delta z_j)\right] 
\end{equation}
where $\Delta z_j = z_j^+ - z_j^-$ is the coordinate difference between the $j$th positive and negative charge pair, $\zeta(z)$ is the Weierstrass zeta function with quasi-periods $2\omega_1$ and $2\omega_2$ \cite{Whittaker_Watson_1996}, and $f(\Delta z_j)$ is a function that only depends on the charges' positions and the center cell geometry, and, hence, may be regarded as a constant for a given configuration.
For a square domain with system length $L$ it reduces to $f(\Delta z_j) = (\pi/L^2)\Delta \bar{z}_j$.
For completeness, a derivation of the right hand side of Eq.~\eqref{eqn:CoulombVelocity} may be found in Appendix \ref{app:Weierstrass}.

A solution to Eq.\eqref{eqn:CoulombVelocity} is given by
\begin{equation} \label{eqn:CoulombOrderParam}
    \Omega(z) = -\sum_{j=1}^N\left[\log\frac{\sigma(z-z_j^+)}{\sigma(z-z_j^-)} + f(\Delta z_j)z\right] + C
\end{equation}
where $\sigma(z)$ is the Weierstrass sigma function [defined by $d\log\sigma/dz = \zeta(z)$] \cite{Whittaker_Watson_1996} and $C$ is an integration constant.
Thus, the imaginary part of Eq.~\eqref{eqn:CoulombOrderParam} represents a candidate analytical solution for the angle of the $p$-atic orientation field.
However, even though its derivative is periodic, Eq.~\eqref{eqn:CoulombOrderParam} is only quasi-periodic:
\begin{equation}
    \Omega(z + 2\omega_k) - \Omega(z) = -\sum_{j=1}^N\left[2f(\Delta z_j)\omega_k - 2\eta_k \Delta z_j\right]
\end{equation}
where $\eta_k = \zeta(\omega_k)$ \cite{Whittaker_Watson_1996}.

For electrostatic charges, quasi-periodicity of the potential is not an issue, as long as its derivative, the electric field, is periodic
For $p$-atics, however, the physically relevant object is the orientation field $\theta_p$, therefore, it must be periodic. 
I show examples of $\theta_p = -(1/p)\im\left[\Omega(z)\right]$ computed from Eq.~\eqref{eqn:CoulombOrderParam} for $p=1$ ($XY$-model) in Fig.~\ref{fig:NonPeriodicConfiguration}(a) and $p=2$ (nematic liquid crystal) in Fig.~\ref{fig:NonPeriodicConfiguration}(b), where it is clear that the periodic boundary conditions are not satisfied.

\subsection{Periodic configuration and interaction between defects}

It follows from the above analysis that, in general, topological defects in $p$-atics do not share the same interaction as Coulomb charges in periodic systems.
What, then, is the interaction between defects?
To answer this, we first need a periodic configuration; fortunately, the imaginary part of Eq.~\eqref{eqn:CoulombOrderParam} can be forced to be periodic by adding terms proportional to $z$.
The angle of the orientation field, $\theta_p(z,\bar{z}) = -(1/p)\im\left[ \Omega(z)\right]$, is then
\begin{multline} \label{eqn:ThetaFinalAnswer}
    \theta_p(z,\bar{z}) = \frac{1}{p}\im\left\{\sum_{j=1}^N\left[\log\frac{\sigma(z-z_j^+)}{\sigma(z-z_j^-)} \right.\right.\\
    \left.\left.+ \left[2i(\eta_1\bar{\omega}_2 - \eta_2\bar{\omega}_1)\Delta z_j - \pi\Delta\bar{z}_j\right]\frac{z}{A}\right] \right. \\ 
    \left. + \frac{2\pi i}{\omega_2 - \bar{\omega}_2}
    \left[-\frac{\bar{\omega}_2}{\omega_1}n_1 + n_2\right]z\right\} +\theta_0
\end{multline}
where $A$ is the domain area, the second to last term comes from the invariance under rotations by $2\pi/p$ ($n_1$ and $n_2$ are integers), and $\theta_0$ is a global phase (all examples shown here will have $\theta_0=0$).
Note that the corresponding analytic potential $\Omega(z)$ will still only be quasi-periodic in general.
Indeed, it is not possible for a function with arbitrary poles to be both doubly periodic and analytic \cite{Whittaker_Watson_1996}.

\begin{figure}
    \centering
    \includegraphics[width = \columnwidth]{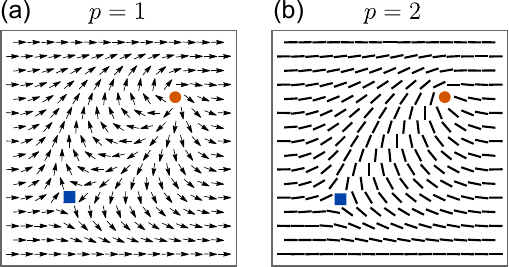}
    \caption{Orientation fields corresponding to Eq.~\eqref{eqn:ThetaFinalAnswer} for the same system of defects as in Figs.~\ref{fig:NonPeriodicConfiguration}(a,b)}
    \label{fig:PeriodicConfiguration}
\end{figure}

The orientation field corresponding to Eq.~\eqref{eqn:ThetaFinalAnswer} is plotted in Figs.~\ref{fig:PeriodicConfiguration}(a,b) for the cases shown previously in Figs.~\ref{fig:NonPeriodicConfiguration}(a,b) (with $n_1 = n_2 = 0$) and it is clear that the boundary conditions are now satisfied.
To predict the defects' velocities we may use Eq.~\eqref{eqn:DefectVelocity}, which gives the velocity of the $k$th positive defect:
\begin{multline} \label{eqn:PositiveDefectVelocity}
    v_k^+ = 2\left\{\sum_{j\neq k^+}\left[\zeta(\bar{z}_k^+-\bar{z}_j^+) - \zeta(\bar{z}_k^+ - \bar{z}_j^-)\right]\right. \\ 
    \left. - \frac{1}{A}\sum_j \left[\pi\Delta z_j + 2i(\bar{\eta}_1\omega_2 - \bar{\eta}_2\omega_1)\Delta \bar{z}_j\right]\right. \\ \left.+\frac{2\pi i}{\omega_2 - \bar{\omega}_2}\left[-\frac{\omega_2}{\bar{\omega}_1}n_1 + n_2\right] \right\}
\end{multline}
where the notation $j\neq k^+$ indicates a sum that excludes only the $k$th positive defect.
The velocity of the negative defects may be found by inverting the sign and excluding the $k$th negative defect from the first sum.
In deriving these equations I have relied on complex-valued functions as the notation is more compact; however, it is possible to derive the same results using only real-valued functions, which I show in Appendix \ref{app:RealValue}.

\begin{figure*}
    \centering
    \includegraphics[width = \textwidth]{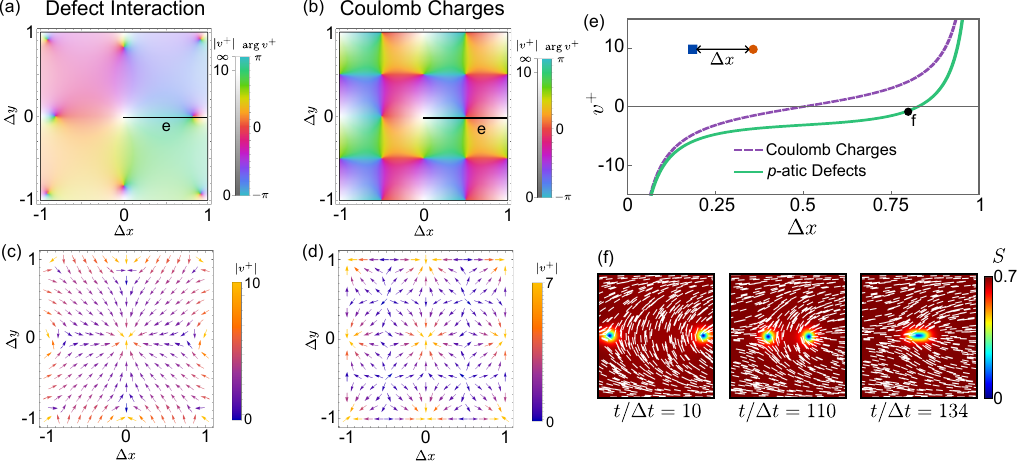}
    \caption{(a,b) Complex color plots of the velocity of the positive $p$-atic defect, Eq.~\eqref{eqn:DefectVel} and force on the positive Coulomb charge, Eq.~\eqref{eqn:CoulombVel} for a configuration with two defects or charges in a square domain with $\omega_1 = 0.5$ and $\omega_2 = 0.5 i$ as a function of defect separation $\Delta z = \Delta x + i\Delta y$.
    The color represents the argument of the function while the brightness indicates the magnitude.
    The black line indicates the subset of the plot that is shown in (e).
    (c,d) Corresponding vector plots of Eqs.~\eqref{eqn:DefectVel} and \eqref{eqn:CoulombVel} where the arrows indicate the argument of the function while the color indicates the magnitude.
    (e) Velocity of the positive defect for the case $\Delta y = 0$ (green solid line).
    The purple dashed line shows the force on the positive Coulomb charge in the same configuration.
    The black dot indicates the initial separation for the simulation shown in (f).
    (f) Time snapshots of a simulation of two annihilating defects in a nematic liquid crystal ($p = 2$) with initial separation $\Delta x = 0.81$.
    The color is the scalar order parameter $S$ and the white lines are the nematic director $\mathbf{\hat{n}}$}
    \label{fig:TwoDefectInteraction}
\end{figure*}

Let us now analyze the predicted defect velocity, Eq.~\eqref{eqn:PositiveDefectVelocity}, for the simple case of two defects in a square domain of length $L = 1$ and $n_1 = n_2 = 0$.
In this case, the velocity simplifies to
\begin{equation} \label{eqn:DefectVel}
    v_{defect}^+ = -2\left[\zeta(\Delta \bar{z}) + \pi \Delta z\right].
\end{equation}
Compare this to the force ($f = f_x +if_y$) experienced by the positive Coulomb charge in the same setup:
\begin{equation} \label{eqn:CoulombVel}
    f_{Coulomb}^+ = -2\left[\zeta(\Delta \bar{z}) - \pi \Delta z\right].
\end{equation}
The two expressions differ only by a sign, yet the sign difference drastically alters the interaction.
In Figs.~\ref{fig:TwoDefectInteraction}(a--d) I have plotted Eqs.~\eqref{eqn:DefectVel} and \eqref{eqn:CoulombVel} as complex color plots and as vector plots.
The color plots, Figs.~\ref{fig:TwoDefectInteraction}(a,b), clearly show the zeros and poles of the interaction.
Poles occur when defects are on top of one another (they are attracted with infinite force) while zeros occur due to force balance through the periodic boundaries.
For electrostatic charges, the zeros are located on a square half the size of the domain since the interaction is symmetric.
If the separation between defects is larger than half the system size, the electrostatic charges will attract through the periodic boundary instead of towards the interior of the domain.
On the other hand, the zeros of the $p$-atic defect interaction are no longer located on a square and occur very close to the periodic boundary.
Additionally, $p$-atic defects will only attract through the periodic boundaries in a small locus of separations near the zeros of the interaction.
The vector plot in Fig.~\ref{fig:TwoDefectInteraction}(c) shows that $p$-atic defects will almost always attract toward the center of the domain.

In Fig.~\ref{fig:TwoDefectInteraction}(e), I have plotted Eqs.~\eqref{eqn:DefectVel} and \eqref{eqn:CoulombVel} for the case where $\Delta y = 0$.
In this one-dimensional cut of Figs.~\ref{fig:TwoDefectInteraction}(a,b) it is easier to compare the two directly.
If the defects are close to each other, either at the center of the domain ($\Delta x = 0$) or at the edge of the domain ($\Delta x = 1$), the defect velocity is similar to the Coulomb interaction.
For two electrostatic charges, the force balance point is $\Delta x = 1/2$ which is required by the symmetry of the interaction.
However, for $p$-atic defects, Eq.~\eqref{eqn:DefectVel} predicts that the force balance point is much closer to the boundary, $\Delta x \approx 0.82$.
I have tested this prediction by simulating the annihilation of two defects in a nematic liquid crystal ($p = 2$) with the initial configuration given by Eq.~\eqref{eqn:ThetaFinalAnswer} with $\Delta x = 0.81$.
The simulation minimizes the Landau-de Gennes free energy \cite{deGennes75} of the nematic order parameter and is subject to periodic boundary conditions (further simulation details are provided in Appendix \ref{app:CompDetails}).
Figure \ref{fig:TwoDefectInteraction}(f) shows several time snapshots of the nematic scalar order parameter $S$ and director $\mathbf{\hat{n}}$, indicating that the defects indeed annihilate in the center even though they are initially closer to each other through the periodic boundary.
In the Supplemental Material, I have provided additional examples of predicted velocities and corresponding simulations for two or more defects that highlight the qualitative difference between defect interactions and Coulomb interactions in PBC \cite{SuppNote25}.

\section{Topological Solitons} \label{sec:Solitons}

\begin{figure}
    \centering
    \includegraphics[width = \columnwidth]{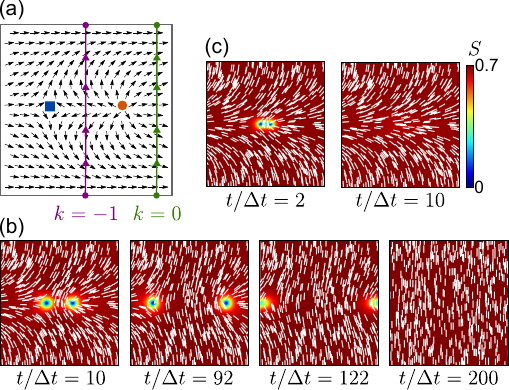}
    \caption{(a) Defects in the $XY$-model subject to periodic boundary conditions.
    The curve between the defects measures a charge [Eq.~\eqref{eqn:DefectCharge}] $k=-1$, indicating a topological excitation.
    If the curve is moved past either defect, it measures $k=0$, as shown by the curve to the right of the positive defect.
    (b) Time snapshots of a nematic liquid crystal simulation with defects initially separated by $\Delta x =0.2$ and $n_2 = 1$.
    Color is the scalar order parameter $S$ while the white lines are the orientation field.
    (c) Time snapshots of a simulation similar to (b) but with initial separation $\Delta x = 0.1$.}
    \label{fig:GlobalSolitons}
\end{figure}

The comparison between Eqs.~\eqref{eqn:DefectVel} and \eqref{eqn:CoulombVel} suggests that one could understand the interaction asymmetry between defects as Coulomb charges with an effective uniform electric field proportional to the distance between defects.
How can we understand the emergence of this effective field physically?
Consider a periodic domain with just two $p$-atic defects and consider a curve $C$ that runs from the bottom to the top of the domain such that the end points are at the same $x$ coordinate.
$C$ is a closed curve in PBC, so we may measure the topological charge, Eq.~\eqref{eqn:DefectCharge}, by this curve.
If we shift the curve past a defect the value of the measured topological charge must change by $\pm1/p$ by definition of a topological defect.
Fig.~\ref{fig:GlobalSolitons}(a) shows an example of this for a defect pair in the $XY$-model ($p = 1$).
The curve between the defects measures a charge $k = -1$, while the curve on the other side of the defects measures a charge $k = 0$.
Such a distortion, which I will call a topological soliton, is required for each defect pair in the system and are stabilized by the periodic boundary condition.
Because topological solitons disrupt the $p$-atic configuration, they cost elastic energy and hence induce additional forces on the defects, thus producing an asymmetric interaction between defect pairs akin to a uniform electric field.

For every pair of $\pm1/p$ defects in the system, there must be a soliton between them.
The sign of the soliton [as measured from the bottom (left) of the domain to the top (right)] depends on the relative position of the defect pair to the measuring curve.
For example, the defect pair shown in Fig.~\ref{fig:GlobalSolitons}(a) has the positive defect to the right of the purple curve and the negative defect to the left which results in the curve measuring $k = -1$.
If the defect positions were switched so that the positive defect was on the left of the measuring curve, one would measure $k = +1$ instead.
The green curve would still measure $k = 0$ since it does not straddle the defects on the interior of the domain.
If there are multiple pairs of defects, the solitons induced by the pairs may either add or subtract to one another, either increasing or reducing the interaction strength between all other defects (a few examples of this are shown in the Supplemental Material \cite{SuppNote25}).
One might worry that this analysis depends on the location of the periodic boundaries since, technically, a periodic domain is not bounded (one should think of a torus or donut).
If I change the location of the ``boundaries'' in Fig.~\ref{fig:GlobalSolitons}(a) so that the green curve straddles the defects, I should still measure $k=0$ since I have only changed my field of view, and not anything within the system.
The resolution to this is that, once one has set the ``boundaries'' of the system, they should be consistently held, and then one has to then take into account an additional, key feature that I have not yet analyzed: the effect of ``global'' solitons.

\subsection{Global Solitons}

Topological solitons may exist without defects.
Indeed, this is the physical interpretation of the integers $n_1$ and $n_2$ in Eqs.~\eqref{eqn:GlobalSolitonAddition} and \eqref{eqn:ThetaFinalAnswer}: In the absence of topological defects, $n_1$ ($n_2$) is the number of windings the orientation field makes along the $\omega_1$ ($\omega_2$) direction, and hence counts the number of ``global'' topological solitons in that direction.
The term ``global'' here refers to the fact that $n_1$ windings in the $\omega_1$ direction occur for all coordinates along $\omega_2$, and vice versa.
When defects are present, these global topological solitons may add or subtract to the local solitons which appear between defect pairs, potentially changing the direction of the interaction.
Figure \ref{fig:GlobalSolitons}(b) shows a simulation of two defects initially separated by distance $\Delta x = 0.2$ with a global soliton added along the $y$-direction, $n_2 = 1$, which has the effect of removing the soliton on the inner domain, and adding a soliton on the outer domain.
For example, compare the first panel of the simulation in Fig.~\ref{fig:TwoDefectInteraction}(f) and the first panel in Fig.~\ref{fig:GlobalSolitons}(b).
The former has a local soliton in the middle of the domain between the defect, while the latter has moved the soliton to the outer boundary.
As the simulation snapshots show, the defects now move to annihilate on the outer boundary, even though they are initially closer to each other inside the domain.

It is not possible for a uniform configuration to spontaneously shift to one with global topological solitons.
It is possible, however, to go from a uniform configuration to a solitonic configuration by nucleation and then subsequent annihilation of a defect pair.
Figure \ref{fig:GlobalSolitons}(c) shows an example of a pair of simulated defects annihilating to stabilize a global soliton.
Here the simulation is initialized in the same way as in Fig.~\ref{fig:GlobalSolitons}(b), except the initial separation is $\Delta x = 0.1$ so the defects are close enough to attract towards the center of the domain.
In general, defects that are interacting via elastic relaxation will move towards the local solitons, unless they are within the locus of points near the zeros of the interaction shown in Fig.~\ref{fig:TwoDefectInteraction}(a).

Transitioning from a uniform configuration to a global soliton costs energy, and so would not occur spontaneously in gradient flow simulations, though could occur in non-equilibrium contexts such as applied external fields or active liquid crystals \cite{giomi13,doo18,Armengol-Collado2023-pk}.
In such systems topological defect unbinding and annihilation events may change the number of global solitons.
To analyze the elastic interaction between defects in these contexts, one needs to keep track of the topological indices $n_1$ and $n_2$ as well as defect positions.
To measure $n_1$ ($n_2$) for an arbitrary configuration one could measure the winding number on a loop that goes along $\omega_2$ ($\omega_1$) and add the winding of the local solitons that are required by the topological defects, as discussed above.
For temporal data, if the indices $n_1$ and $n_2$ are known at a given time, the following rule may be used to keep track of $n_1$ and $n_2$ for future times:
If a positive defect moves through the periodic boundary in the positive $\omega_2$ direction, then $n_1\to n_1 + 1$ and if a positive defect moves through the periodic boundary in the positive $\omega_1$ direction, then $n_2 \to n_2 - 1$.
The rule for the negative direction is the opposite, and rules for negative defects apply opposite to those for positive defects.
Additionally, if one also wants to keep track of the configuration $\theta_p$, then every time a defect passes a periodic boundary one should add $\pi/p$ to $\theta_0$.
This will ensure that the orientation of defects remain continuous as they move across the periodic boundary.
Considering the effects of global solitons, it is interesting to note that the defect behaviors not only depend on their current positions, but also on the history of the defect dynamics in the system.

\section{Physically Realizable Domains} \label{sec:Annulus}

\begin{figure}
    \centering
    \includegraphics[width = \columnwidth]{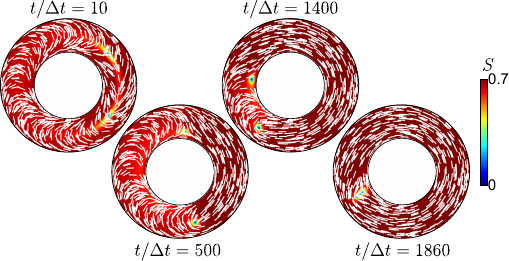}
    \caption{Time snapshots of a nematic liquid crystal simulation in an annulus with radii $r_1 = 1,\,r_2=2$ and initial angular defect separation $\Delta \phi = 2\pi/5$ with the local topological soliton on the longer side of the annulus.
    Color is the scalar order parameter $S$ while the white lines are the orientation field.}
    \label{fig:AnnulusSimulation}
\end{figure}

Since PBC are reserved to theoretical or computational studies, a natural question is whether these effects would be observable in experimentally relevant domains.
One case where this may be observed is in annular geometries such that the boundary conditions on the inner and outer boundary are the same.
The argument for the existence of solitons between defects still applies, and configurations from a square geometry with periodic boundaries on one edge may be mapped to the annulus.
In Fig.~\ref{fig:AnnulusSimulation} I show simulation snapshots of a nematic liquid crystal in an annular geometry (radii $r_1 =1$ and $r_2 = 2$) with two defects initialized with angular separation $\Delta \phi = 2\pi/5$ but such that the topological soliton between them is on the longer side of the domain.
In this case the defects annihilate along the longer path even though they are initially close to each other in the annulus.
Note that here the defect trajectories are not symmetric due to the influence of the boundary curvature which interacts differently with positive and negative defects.
Even without a quantitative prediction for the velocity we can estimate the direction the defects will move based on the location of the topological solitons between them.

Another geometry with these properties that may be physically realized is a torus, since this can be mapped directly to a periodic domain.
The concepts of local and global topological solitons would apply to the torus, so it may be possible to observe interactions similar to those shown in Fig.~\ref{fig:AnnulusSimulation} for the annulus where defects annihilate over the longer part of the domain.
Simulations on the torus are beyond the scope of this paper, and I note that curvature plays an important role in this system as well, since it is known that Gaussian curvature attracts defects of like sign, while extrinsic curvature induces a preferred orientation \cite{Selinger2011-gm,ellis18,Rojo-Gonzalez24}, so it is unclear what the relative strength of the topological forces may be.

\section{Conclusion} \label{sec:Conclusion}

The effect of boundary conditions on phases with broken rotational symmetry has been of great interest for decades.
While fixed (Dirichlet) boundary conditions are often the subject of such studies, I have shown analytically that PBC bestow a different interaction between defects than in infinite domains.
The interaction is due to the domain topology, requiring non-singular elastic excitations that must accompany defect pairs.
I stress that the defect interactions are qualitatively different than those of Coulomb charges, and that this should be recognized when comparing computational data to experimental data.
Further, additional topological indices ($n_1$ and $n_2$) which represent overall windings along the periodic lattice directions, must be taken into account when considering the elastic forces on defects, and these windings depend on the history of the defect dynamics.

A future challenge will be to adapt previous work regarding arbitrary defect orientations \cite{vromans16,tang17,pearce21,Copar24} to the case of periodic bounded systems.
Here I have assumed that the relative orientations between defects have the lowest energy.
It would also be interesting to test the predictions on toroidal geometries where curvature will also play a role in defect dynamics and to posit topological interactions for more complex domains such as the $n$-genus surface.
Additionally, generalizing these results to three dimensions will be important for analyzing the interactions between defects in three-dimensional ferromagnets or nematics, which no longer share the same topology and types of defects.
The methods and results of this work will prove useful in these future endeavors.

\begin{acknowledgements}
I am indebted to Oleg Tchernyshyov for pointing out the connection between the orientation field and the electrostatic potential, as well as for the use of the Weierstrass functions to describe the electric field of periodic point charges. This work also benefited greatly from conversations with Brandon Klein and Daniel Beller. I gratefully acknowledge support from the Department of Physics and Astronomy at Johns Hopkins University through the William H. Miller III postdoctoral fellowship.
\end{acknowledgements}

\appendix

\section{Halperin-Mazenko formalism for $p$-atic defect velocities} \label{app:HalperinMazenko}
For completeness, I derive here the equation for the velocity of defects in a $p$-atic liquid crystal using the formalism of Halperin and Mazenko \cite{halperin81,liu92,mazenko97,mazenko99}.
This will mirror the derivation for active nematic liquid crystals in Ref. \cite{angheluta21}, except that in this case the order parameter has $p$-fold rotational symmetry and the only time dependence I will consider comes from the elastic interaction.

The $p$-atic order parameter may be written as a complex-valued function
\begin{equation} 
    \psi_p(z,\zb) = S_p(z,\zb)e^{ip\theta_p(z,\zb)}
\end{equation}
where $S_p$ is the local degree of orientational order and $\theta_p$ is the local orientational angle.
As a side note, the reader should not confuse the order parameter with the local director $n_p(z,\zb) = e^{i\theta_p}$ which gives the local orientation of the $p$-atic phase and is implicitly understood to be $p$-fold degenerate under rotations by $2\pi/p$.
$\psi_p$ is defined so that it is explicitly invariant under $\theta_p \to \theta_p + 2\pi/p$.

The magnitude of the order parameter $S_p$ is constant and $S_p \approx 1$ when changes in $\theta_p$ are small, but $S_p\to0$ if $\theta_p$ becomes discontinuous to preserve the continuity of the order parameter.
This is precisely the case for the core of topological defects, and the Halperin-Mazenko formalism takes advantage of this fact.
The topological defect density may be written in two ways:
\begin{equation} \label{Seqn:DefectDensity} 
    \rho_p(z,\zb) = \sum_j k_j\delta(z -z _j)=\frac{1}{p}D_p(z,\zb)\delta\left[\psi_p(z,\zb)\right]
\end{equation}
where $j$ indexes individual defects and $k_j$ is the charge of the $j$th defect.
The first equality is the usual definition for defect charge density while the second equality comes from a coordinate change from the domain to order parameter space.
The Jacobian determinant $D_p$ is given by
\begin{equation} \label{Seqn:DDef} 
    D_p(z,\zb) = \partial_z\psi_p\partial_{\zb}\bar{\psi_p} - \partial_z \bar{\psi_p}\partial_{\zb}\psi_p
\end{equation}
where $\partial_z = (1/2)(\partial/\partial x - i \partial/\partial y)$ and $\partial_{\zb} =\bar{\partial_z}$.
To get the defect velocity, one takes advantage of two conservation laws:
\begin{align}
    \partial_t\rho_p + \nabla \cdot J^{(\rho)} &= 0  \\
    \partial_tD_p + \nabla \cdot J_p^{(\psi)} &=  0 
\end{align}
where the $J$ are respective currents.
The first comes from topological charge conservation, the second can be derived from taking a time derivative of Eq.~\eqref{Seqn:DDef}.
Using the conservation laws and Eq.~\eqref{Seqn:DefectDensity}
\begin{equation} 
    J^{(\rho)} = \sum_jv_jk_j\delta(z-z_j) = \sum_j \frac{J_p^{(\psi)}}{D_p}k_j\delta(z-z_j)
\end{equation}
where the first equality can be derived by taking a time derivative of Eq.~\eqref{Seqn:DefectDensity} and defining $v_j = \partial z_j/\partial t$ and the second equality combines the conservation of $D_p$ with the definition of the defect density.
Comparing the equations, and noting that the $\delta$-functions require this only hold at $z = z_j$, yields
\begin{equation} \label{Seqn:OPVelocity} 
    v_j =\frac{J_p^{(\psi)}}{D_p}= \left.\frac{-\partial_t\psi_p\partial_{\zb}\bar{\psi_p} + \partial_t\bar{\psi_p}\partial_{\zb}\psi_p}{\partial_z\psi_p\partial_{\zb}\bar{\psi_p} - \partial_z \bar{\psi_p}\partial_{\zb}\psi_p}\right|_{z=z_j}
\end{equation}
where $\partial_t = \partial/\partial t$.

Equation \eqref{Seqn:OPVelocity} gives a kinematic equation for the velocity of a defect as a function of the local order parameter.
To get a prediction for the defect velocity one must approximate both the space and time derivatives of $\psi_p$.
To handle the time derivative, I will assume a Landau-Ginzburg model for the free energy of the $p$-atic phase:
\begin{equation} \label{Seqn:ComplexFreeEnergy} 
    F_p = \int\left[\frac{A}{2}|\psi_p|^2 + \frac{C}{4}|\psi_p|^4 + \frac{K}{2}|\nabla\psi_p|^2\right]\,d\mathbf{r}
\end{equation}
where $A$ and $C$ are phenomenological parameters (set so that the ordered phase is stable) and $K$ is the elastic constant. 
The time dependence of the order parameter is further assumed to be given by free energy relaxation
\begin{equation} \label{Seqn:Relaxation} 
    \partial_t\psi_p = -\frac{1}{\gamma}\frac{\delta F}{\delta\psi_p}=-\frac{1}{\gamma}\left[\left(A + C|\psi_p|^2\right)\psi_p - 4K\partial_z\partial_{\zb}\psi_p\right]
\end{equation}
where $\gamma$ is a rotational viscosity. 

Equation \eqref{Seqn:Relaxation} can be substituted into Eq.~\eqref{Seqn:OPVelocity} to yield an expression that depends only on spatial derivatives of $\psi_p$ (terms proportional to $\psi_p$ go to zero at the defect core).
To handle the spatial derivatives I use the fact that near the defect core $S_p \sim |z-z_j|$ so that 
\begin{equation}  
    \psi_p(z,\zb) \approx |z - z_j|\left(\frac{z-z_j}{\zb - \zb_j}\right)^{\pm\frac{1}{2}}e^{ip\tilde{\theta}_p(z,\zb)}
\end{equation}
where the $+$ $(-)$ is used for a positive (negative) defect and $\tilde{\theta}_p$ is the non-singular orientation field (i.e. the orientation field due to all other defects).
Substituting this into Eq.~\eqref{Seqn:OPVelocity} yields Eq.~\eqref{eqn:DefectVelocity} above:
\begin{equation} \label{Seqn:DefectVelocity} 
    v_j = -\frac{4p^2k_jK}{\gamma}i\partial_{\zb}\tilde{\theta}_p.
\end{equation}

It is interesting to note that since the charges are $\pm1/p$ the dependence of $p$ drops out of the velocity, predicting that defects elastically interact at the same strength independent of $p$.
This is initially surprising, since the energy of a defected configuration is smaller for larger $p$ (the field rotates by less, hence the overall gradient is smaller), so the rotation rate of the relaxing $p$-atic phase should be smaller.
However, for increasing $p$, the required rotation to move a defect a given distance is smaller, exactly accounting for the slower rotation rate.

\section{The Coulomb interaction and Weierstrass functions} \label{app:Weierstrass}
Equation \eqref{eqn:CoulombVelocity} of the main text asserts that the velocity of defects that interact as Coulomb charges in a periodic domain will be given by Weierstrass $\zeta$ functions and an additional constant that depends on the domain shape and size and the locations of defects.
I show this here for completeness.

The electric field of a point charge in two-dimensions may be written in complex coordinates as
\begin{equation} 
    E = E_x - iE_y = \frac{1}{z -z_0}
\end{equation}
where $z_0$ is the location of the point charge and I am using units so that $e/2\pi\epsilon_0 = 1$.
In the periodic configuration described in the main text, with half periods $\omega_1$ and $\omega_2$, the electric field will be the superposition of all fields created by the periodic copies:
\begin{equation} \label{Seqn:EFieldSummation} 
    E = \frac{1}{z-z_0} + \sum_{\ell,m\neq0}\frac{1}{z - z_0 - 2\omega_1\ell - 2\omega_2m}.
\end{equation}
Compare this with the definition of the Weierstrass zeta function \cite{Whittaker_Watson_1996}:
\begin{equation} 
    \zeta(z) =\frac{1}{z} +\sum_{\ell,m\neq0}\left[\frac{1}{z - \Omega_{\ell,m}} + \frac{1}{\Omega_{\ell,m}}+\frac{z}{\Omega_{\ell,m}^2}\right]
\end{equation}
where $\Omega_{\ell,m} = 2\omega_1\ell + 2\omega_2m$.
Thus, evidently, Eq.~\eqref{Seqn:EFieldSummation} may be written as
\begin{equation} \label{Seqn:EFieldNotPeriodic} 
    E = \zeta(z - z_0) - \sum_{\ell,m\neq0}\left[\frac{1}{\Omega_{\ell,m}} + \frac{z-z_0}{\Omega_{\ell,m}^2}\right].
\end{equation}
There is, however, an issue with Eq.~\eqref{Seqn:EFieldNotPeriodic}: it is only quasi-periodic.
The quasi-periodicity is a result of the distribution of fields for a periodic array of charges \cite{Hill21}.
A periodic function can be constructed by adding terms proportional to $z - z_0$ and $\zb - \zb_0$ \cite{Hill21}.
The result is
\begin{multline} \label{Seqn:EFieldPeriodic} 
    E(z,\zb) = \zeta(z-z_0) -\frac{2i}{A}\left(\eta_1\bar\omega_2-\eta_2\bar\omega_1\right)(z-z_0) \\ 
    -\frac{\pi}{A}(\zb - \zb_0)- \sum_{\ell,m\neq0}\frac{1}{\Omega_{\ell,m}}
\end{multline}
where $A$ is the area of the domain.
One may worry that the field in Eq.~\eqref{Seqn:EFieldPeriodic} is no longer analytic, and hence does not obey Maxwell's equations for a single point charge.
The resolution is that the non-analytic part (proportional to $\zb$) actually adds a constant uniform density electric charge that balances the point charge.
This is consistent with the requirement of zero net electric flux through the periodic boundaries, i.e. the total charge on a torus must be zero.

Finally, to yield Eq.~\eqref{eqn:CoulombVelocity} in the main text, I note that the electric field produced by two charges of opposite sign is
\begin{align}
    E &= \zeta(z-z^+) - \zeta(z-z^-) + f(\Delta z)  \\
    f(\Delta z) &= \frac{2i}{A}\left(\eta_1\bar\omega_2 - \eta_2\bar\omega_1\right)\Delta z + \frac{\pi}{A}\Delta\zb 
\end{align}
which can be shown from Eq.~\eqref{Seqn:EFieldPeriodic}.
Since the charge is balanced in this configuration, the resulting electric field is analytic.

\section{Real-valued orientation configuration and defect velocity} \label{app:RealValue}
The arguments in the main text using complex-valued functions can be made using only real-valued functions.
I derive the resulting expressions here.
Equation \eqref{eqn:DefectVelocity} can be equivalently derived for real-valued order parameters (see e.g. Ref. \cite{schimming23}).
For $p$-atics it takes the form
\begin{equation} \label{Seqn:RealDefectVelocity} 
    \mathbf{v}_j=\frac{2p^2k_j K}{\gamma}\left.\left(\frac{\partial \tilde{\theta}_p}{\partial_y},\,-\frac{\partial\tilde{\theta}_p}{\partial x}\right)\right|_{(x_j,y_j)}
\end{equation}
where $(x_j,y_j)$ is the location of the $j$th defect.

\begin{widetext}
As in the main text, I will first assume that the interaction is that of Coulomb charges, so that the orientation field should satisfy
\begin{align}
    \frac{\partial\theta_p}{\partial y}&=\sum_{i=1}^N\sum_{\ell,m\in\mathbb{Z}}\left[\frac{x-x_i^+-\ell L_x}{(x -x_i^+ - \ell L_x)^2 + (y-y_i^+-mL_y)^2}-\frac{x-x_i^--\ell L_x}{(x -x_i^- - \ell L_x)^2 + (y-y_i^--mL_y)^2}\right] \label{Seqn:RealCoulombx} \\
    -\frac{\partial\theta_p}{\partial x}&=\sum_{j=1}^N\sum_{\ell,m\in\mathbb{Z}}\left[\frac{y-y_i^+-m L_y}{(x -x_i^+ - \ell L_x)^2 + (y-y_i^+-mL_y)^2}-\frac{y-y_i^--m L_y}{(x -x_i^- - \ell L_x)^2 + (y-y_i^--mL_y)^2}\right] \label{Seqn:RealCoulomby}
\end{align}
where $i$ indexes the defect pair and $\mathbb{Z}$ denotes the set of integers.
Here I have restricted the domain to be rectangular for simplicity; the following can be easily generalized to arbitrary parallelogram domains \cite{Gronbech-Jensen99}.
Equations \eqref{Seqn:RealCoulombx} and \eqref{Seqn:RealCoulomby} can be simplified by summing on one of the indices first using a method from the computational electrostatics literature known as Lekner summation \cite{Lekner91,Gronbech-Jensen96,Gronbech-Jensen99}.
Summing on $\ell$ and then integrating the result gives
\begin{multline} \label{Seqn:RealThetaNotPeriodic} 
    \theta_p(x,y) = \sum_{i=1}^N\left\{-\frac{1}{p}\sum_{m\in\mathbb{Z}}\Bigg[\Arctan\left[\coth\frac{\pi}{L_y}(y-y_i^+ +mL_y)\tan\frac{\pi}{L_x}(x-x_i^+)\right]\right. \\
    \left.- \Arctan\left[\coth\frac{\pi}{L_y}(y-y_i^- +mL_y)\tan\frac{\pi}{L_x}(x-x_i^-)\right]\Bigg]+\frac{2\pi}{p}\frac{\Delta y_i x}{L_x L_y}\right\}
\end{multline}
where $\Delta y_i = y_i^+ - y_i^-$.

Equation \eqref{Seqn:RealThetaNotPeriodic} is equivalent to Eq.~\eqref{eqn:CoulombOrderParam} in the main text and still has the problem that it is only quasi-periodic.
Indeed, it satisfies
\begin{equation} 
    \theta_p(x+\ell L_x,y+mL_y) - \theta_p(x,y) = \frac{2\pi}{p}\sum_i^N\left[\Delta y_i \ell + \Delta x_i m\right].
\end{equation}
As with the complex-valued version, we only have to slightly modify Eq.~\eqref{Seqn:RealThetaNotPeriodic} to force periodicity: $\theta_p \to \theta_p - (2\pi/p)\sum_i\left[\Delta y_i x + \Delta x_i y\right]$.
The full expression for the orientation angle is
\begin{multline} \label{Seqn:RealThetaPeriodic} 
    \theta_p(x,y) = \sum_{i=1}^N\left\{-\frac{1}{p}\sum_{m\in\mathbb{Z}}\Bigg[\Arctan\left[\coth\frac{\pi}{L_y}(y-y_i^+ +mL_y)\tan\frac{\pi}{L_x}(x-x_i^+)\right]\right. \\
    \left.- \Arctan\left[\coth\frac{\pi}{L_y}(y-y_i^- +mL_y)\tan\frac{\pi}{L_x}(x-x_i^-)\right]\Bigg]-\frac{2\pi}{p}\frac{\Delta x_i y}{L_x L_y}\right\} + \frac{2\pi}{p}\left(\frac{n_1 x}{L_x} + \frac{n_2 y}{L_y}\right) + \theta_0
\end{multline}
where $\theta_0$ is a global phase and the second to last term comes from the invariance under rotations by $2\pi/p$ ($n_1$ and $n_2$ are integers).
Equation \eqref{Seqn:RealThetaPeriodic} is the real-valued equivalent of Eq.~\eqref{eqn:ThetaFinalAnswer} in the main text (when $\omega_2/\omega_1 = iL_y/L_x$).
I note that there is another equivalent expression if $m$ is summed over in Eqs.~\eqref{Seqn:RealCoulombx} and \eqref{Seqn:RealCoulomby}:
\begin{multline} \label{Seqn:RealThetaPeriodicAlt} 
    \theta_p(x,y) = \sum_{i=1}^N\left\{\frac{1}{p}\sum_{\ell\in\mathbb{Z}}\Bigg[\Arctan\left[\coth\frac{\pi}{L_x}(x-x_i^+ +\ell L_x)\tan\frac{\pi}{L_y}(y-y_i^+)\right]\right. \\
    \left.- \Arctan\left[\coth\frac{\pi}{L_x}(x-x_i^- +\ell L_x)\tan\frac{\pi}{L_y}(y-y_i^-)\right]\Bigg]-\frac{2\pi}{p}\frac{\Delta y_i x}{L_x L_y}\right\} + \frac{2\pi}{p}\left(\frac{n_1 x}{L_x} + \frac{n_2 y}{L_y}\right) + \theta_0.
\end{multline}

The velocity for a positive defect can be obtained by substituting either Eq.~\eqref{Seqn:RealThetaPeriodic} or Eq.~\eqref{Seqn:RealThetaPeriodicAlt} into Eq.~\eqref{Seqn:RealDefectVelocity}.
Using Eq.~\eqref{Seqn:RealThetaPeriodic},
\begin{multline} \label{Seqn:RealDefectXVelocity} 
    v^+_{x,j} = \frac{2K}{\gamma}\left\{\sum_{i\neq j^+}\sum_{m\in\mathbb{Z}}\frac{\pi}{L_y}\left[-\frac{\sin \frac{2\pi}{L_x}(x_j^+ - x_i^+)}{\cos\frac{2\pi}{L_x}(x_j^+ - x_i^+)-\cosh\frac{2\pi}{L_y}(y_j^+ - y_i^+ + mL_y)} \right.\right. \\
    \left.\left.+ \frac{\sin \frac{2\pi}{L_x}(x_j^+ - x_i^-)}{\cos\frac{2\pi}{L_x}(x_j^+ - x_i^-)-\cosh\frac{2\pi}{L_y}(y_j^+ - y_i^- + mL_y)}\right] +\frac{2\pi}{L_y}\left[n_2 - \sum_i\frac{\Delta x_i}{L_x}\right]\right\}
\end{multline}
\begin{multline} \label{Seqn:RealDefectYVelocity} 
    v^+_{y,j} = \frac{2K}{\gamma}\left\{\sum_{i\neq j^+}\sum_{m\in\mathbb{Z}}\frac{\pi}{L_x}\left[-\frac{\sinh \frac{2\pi}{L_y}(y_j^+ - y_i^+ + mL_y)}{\cos\frac{2\pi}{L_x}(x_j^+ - x_i^+)-\cosh\frac{2\pi}{L_y}(y_j^+ - y_i^+ + mL_y)} \right.\right. \\
    \left.\left.+ \frac{\sinh \frac{2\pi}{L_y}(y_j^+ - y_i^- + mL_y)}{\cos\frac{2\pi}{L_x}(x_j^+ - x_i^-)-\cosh\frac{2\pi}{L_y}(y_j^+ - y_i^- + mL_y)}\right] - \frac{2\pi}{L_x}n_1\right\}
\end{multline}
where $v_{x,j}^+$ denotes the $x$ component of the velocity of the $i$th positive defect and the notation $i \neq j^+$ indicates summation on all defect positions that are not the $i$th positive defect.
Equations \eqref{Seqn:RealDefectXVelocity} and \eqref{Seqn:RealDefectYVelocity} are the real-valued equivalents of Eq.~(8) in the main text.
\end{widetext}

Finally, even though Eqs.~\eqref{Seqn:RealThetaPeriodic}--\eqref{Seqn:RealDefectYVelocity} include infinite sums, they are actually quickly converging sums due to the hyperbolic trigonometric functions.
The infinite sums appearing in the equations will converge to machine precision [error $\sim O(10^{-16})$] after summing up to $|m|,|\ell|=5$ \cite{Gronbech-Jensen96}.
I have provided a simple MATLAB script which computes the orientation angle $\theta_p(x,y)$ and the predicted velocities of defects given the positions of defects to facilitate future studies \cite{PeriodicDefectCode25}.

\section{Computational details} \label{app:CompDetails}
Here I give details regarding the simulations described in the main text and Supplemental Material.
The simulations solve the time evolution equations for the nematic tensor order parameter $\mathbf{Q} = S[\mathbf{\hat{n}}\otimes\mathbf{\hat{n}} - (1/2)\mathbf{I}]$ where $S$ is the local degree of order and $\mathbf{\hat{n}}$ is the director.
$\mathbf{Q}$ may be mapped to the complex order parameter by $S_2 e^{2i\theta_2}=\sqrt{2}(Q_{11} + iQ_{12})$, hence the procedure described here may also be regarded as a simulation of the complex-valued order parameter.

The free energy of a nematic configuration is given by the Landau-de Gennes free energy \cite{deGennes75}:
\begin{equation}
    F = \int \left[\frac{A}{2}|\mathbf{Q}|^2 + \frac{C}{4}|\mathbf{Q}|^4 + \frac{K}{2}|\nabla \mathbf{Q}|^2|\right]\, d\mathbf{r}
\end{equation}
and for all simulations I set $K = 10^{-3}$, $C = 4$, and $A = -1$ which sets the value of $S$ in the ordered phase to $S^*=1/\sqrt{2}$.
To minimize the free energy for a given starting configuration, the time dependence of $\mathbf{Q}$ is
\begin{equation} \label{Seqn:QEvolution}
    \partial_t\mathbf{Q} = -\frac{1}{\gamma}\frac{\delta F}{\delta\mathbf{Q}}=-\frac{1}{\gamma}\left[(A + C|\mathbf{Q}|^2)\mathbf{Q} - K\nabla^2\mathbf{Q}\right]
\end{equation}
where $\gamma$ is a rotational viscosity that is set to $\gamma = 1$ for all simulations.

Equation \eqref{Seqn:QEvolution} is discretized in space on a square mesh of length $L = 1$ with $20,004\text{--}29,041$ vertices or an annular mesh with radii $r_1 = 1$ and $r_2 =2$ with $20,004$ vertices using the MATLAB/C++ finite element package FELICITY \cite{walker18}.
The time dependence is solved using a backwards Euler method with time step $\Delta t = 0.5$ and a Newton-Raphson sub-routine to solve the resulting nonlinear equation.
In the square domain, periodic boundary conditions are used while in the annular domain Dirichlet boundary conditions such that $S = S^*$ and $\theta_p = \phi$ (where $\phi$ is the azimuthal angle) are used on the inner and outer boundary.
Defect configurations are initialized using the periodic configuration Eq.~\eqref{Seqn:RealThetaPeriodic} with $p=2$.

\bibliography{LC}

\end{document}

\title{Supplemental Material for ``Defect Interactions Through Periodic Boundaries in Two-Dimensional $p$-atics''}

\author{Cody D. Schimming}
\email[]{cschimm2@jh.edu}
\affiliation{Department of Physics and Astronomy, Johns Hopkins University, Baltimore, Maryland, 21211, USA}

\maketitle

\section{Example defect interactions}
Here I supplement the results of the main text with several examples of anomalous defect behaviors when compared to the forces on Coulomb charges in periodic boundary conditions.
I note that some of these examples may not be energetically favorable; however, it is likely that such configurations could occur in non-equilibrium settings such as active nematics, or even in simulated quenches from an isotropic state.
The specific predicted behaviors also have Supplemental Movies accompanying them, captions for which are included in the next section.

\subsection{One Pair}

For a single pair of defects in a square domain ($L_x = L_y = 1$) with $n_1=n_2=0$ the velocity of the positive defect is
\begin{equation} \label{Seqn:2DefectVelocity} \tag{S1}
    v_{defect}^+ = -2[\zeta(\Delta \zb) +\pi\Delta z].
\end{equation}
On the other hand, the force on a positive Coulomb charge interacting with a negative charge in the same domain is
\begin{equation} \label{Seqn:2CoulombVelocity} \tag{S2}
    f_{Coulomb}^+ = -2[\zeta(\Delta \zb) - \pi\Delta z].
\end{equation}

\renewcommand{\thefigure}{S1}
\begin{figure}
    \centering
    \includegraphics[width = \textwidth]{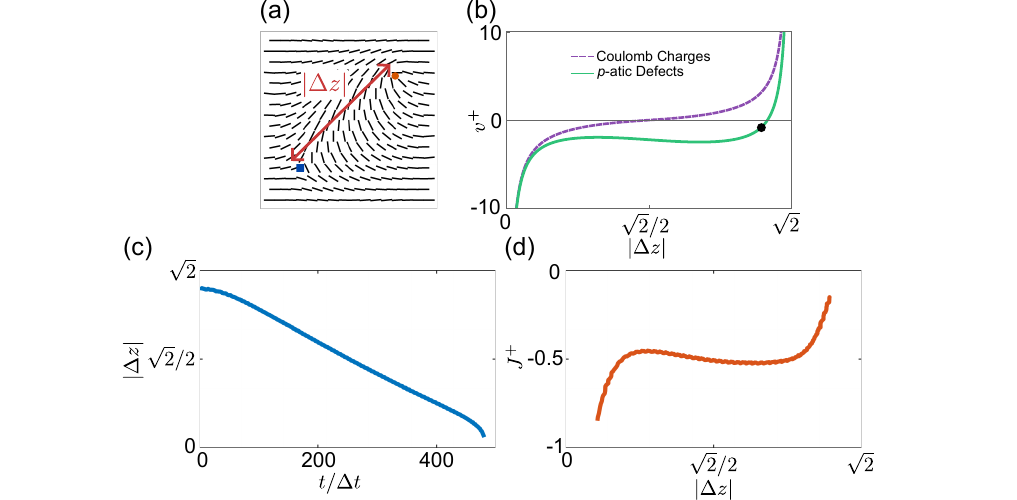}
    \caption{(a) Configuration of nematic defects ($p=2$) located at distance $|\Delta z|$ along the diagonal of the domain.
    (b) Predicted velocity of positive defect and positive Coulomb charge as a function of $|\Delta z|$ for the configuration in (a).
    The black dot indicates the initial condition for the simulation analyzed in (c,d) and shown in Supplemental Movie 5.
    (c) $|\Delta z|$ versus time step $t / \Delta t$ for simulated nematic defects with initial separation $|\Delta z| = 0.9\sqrt{2}$.
    (d) Topological current at the positive defect $J^+$ as a function of defect separation $|\Delta z|$ for simulated nematic defects.
    The current is computed from derivatives of the order parameter via Eq.~(A7) in Appendix A and is proportional to the defect velocity.}
    \label{Sfig:DiagonalDefects}
\end{figure}

An interesting case is when defects are aligned along the diagonal of the domain, as shown in Fig.~\ref{Sfig:DiagonalDefects}(a).
In this case, the predicted velocity of the positive defect is even more asymmetric. 
This is plotted in Fig.~\ref{Sfig:DiagonalDefects}(b) along with the respective force on Coulomb charges to compare.
For $p$-atic defects, the prediction is that the velocity will even be non-monotonic as a function of $|\Delta z|$.
To test this prediction, I simulated defects with an initial $|\Delta z| = 0.9\sqrt{2}$ (Supplemental Movie 5).
Figure \ref{Sfig:DiagonalDefects}(c) plots $|\Delta z|$ versus the simulation time step $t /\Delta t$ until the defects annihilate.
Even though the defects are initially located very close to the outer boundary (i.e. very close to each other), they still annihilate in the interior of the domain.
It is hard to see from Fig.~\ref{Sfig:DiagonalDefects}(c) whether the velocity is non-monotonic or not, so in Fig.~\ref{Sfig:DiagonalDefects}(d) I plot the topological current $J$ at the positive defect computed directly from the order parameter [see Eq.~(A7) in the appendix] versus $|\Delta z|$.
Figure \ref{Sfig:DiagonalDefects} shows a non-monotonic topological current just as predicted by Eq.~\eqref{Seqn:2DefectVelocity}.

\renewcommand{\thefigure}{S2}
\begin{figure}
    \centering
    \includegraphics[width = \textwidth]{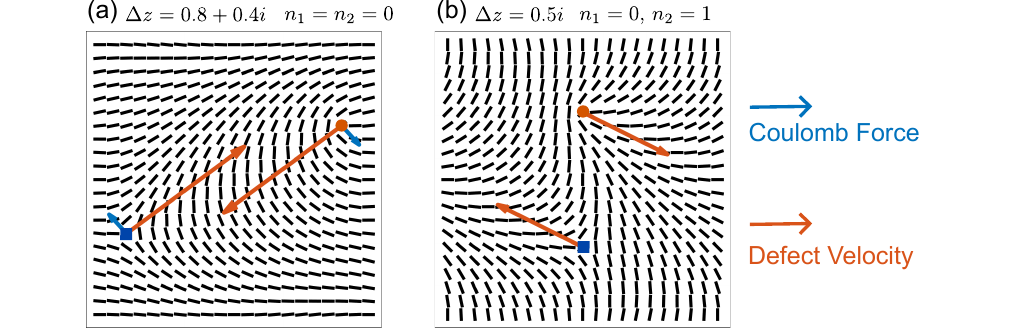}
    \caption{(a) Nematic configuration and predicted velocities for a system with two defects at separation $\Delta z = 0.8 +0.4i$ and $n_1 = n_2 = 0$.
    The orange arrows show the predicted defect velocities while the blue arrows show the Coulomb forces that charges located at the same positions would experience.
    The arrows showing the predicted defect velocities and Coulomb force have the same scale factor.
    (b) Nematic configuration and predicted velocities for a system with two defects at separation $\Delta z = 0.5 i$ and $n_1 = 0$, $n_2 = 1$.
    Here the force experienced by Coulomb charges would be zero.}
    \label{Sfig:TwoDefectInteractionExamples}
\end{figure}

If defects are not aligned along the $x$ or $y$ directions, or the diagonal, the predicted velocity of the two defects is not even co-axial with the Coulomb force on charges at the same location.
An example of this is shown in Fig.~\ref{Sfig:TwoDefectInteractionExamples}(a) for two defects with separation $\Delta z = 0.8 + 0.4i$.
The predicted velocity is toward the center of the domain, even though the force that Coulomb charges would experience would be nearly orthogonal and toward the outer boundary.
Also note that the magnitude of the velocity is nearly five times that of the force between Coulomb charges at the same locations.
Supplemental Movie 6 shows that nematic defects indeed move in the predicted direction when simulated.

Finally, there are interesting effects when $n_1$ or $n_2$ are nonzero.
As an example, Fig.~\ref{Sfig:TwoDefectInteractionExamples}(b) shows defects with separation $\Delta z = 0.5 i$ but also $n_2 = 1$.
The effect of adding the global soliton along the $y$-direction is like adding an external field in the $x$-direction, hence the defects' predicted velocities pick up large $x$-components.
The simulation shown in Supplemental Movie 7 shows that the defects move in the predicted direction.
I note that the force that Coulomb charges experience would be zero in this configuration.

\subsection{Two pairs}

\renewcommand{\thefigure}{S3}
\begin{figure}
    \centering
    \includegraphics[width = \textwidth]{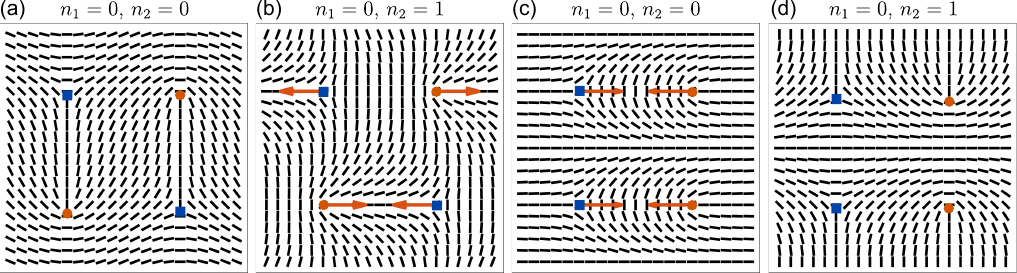}
    \caption{Nematic configurations for four defects in various arrangements.
    (a,b) Positive defects located at $z^+ = 0.25 +0.25i,\,-0.25-0.25i$ and negative defects located at $z^- = 0.25-0.25i,\,-0.25+0.25i$ with global solitons $n_1 = 0$ and $n_2 = 0$ (a) or $n_2 = 1$ (b).
    (c,d) Positive defects located at $z^+ = 0.25 + 0.25i,\,0.25-0.25i$ and negative defects located at $z^- = -0.25 + 0.25i,\,-0.25-0.25i$ with global solitons $n_1 = 0$ and $n_2 = 0$ (c) or $n_2 = 1$ (d).
    The arrows indicate the predicted defect velocities.
    The Coulomb force experienced by charges in each of these configurations would be zero.}
    \label{Sfig:FourDefectInteractionExamples}
\end{figure}

The case of multiple defects is much harder to analyze systematically.
Interestingly, there is at least one specific case in which defects are predicted to behave exactly as Coulomb charges: $\sum_j \Delta z_j =0$.
This condition cannot be met for just two charges, but amounts to the condition that the vector sum of positive defect positions is equal to the vector sum of negative defect positions.
I note that this is also the condition on the location of zeros and poles of a doubly periodic complex-valued function to be represented purely by Weierstrass $\sigma$-functions \cite{Whittaker_Watson_1996}.
An example of this is shown in Fig.~\ref{Sfig:FourDefectInteractionExamples}(a) where positive defects are located at $z^+ = 0.25 +0.25i,\,-0.25 -0.25i$ and negative defects are located at $z^- = 0.25 -0.25i,\, -0.25+0.25i$ and $n_1 = n_2 = 0$.
In this case, the predicted velocity is zero for all defects and this would also be the Coulomb force experienced by electrostatic charges in the same configuration.
Supplemental Movie 8 shows that defects in the simulation do not move in this configuration.
We can understand the lack of motion of defects by counting the solitons between them.
On either side of defects, and in either direction, there are no solitons and hence no distinct direction that the defects should move to lower the energy.

By adding a global soliton, motion of the defects can be induced.
Figure \ref{Sfig:FourDefectInteractionExamples}(b) shows the same configuration of defects as in Fig.~\ref{Sfig:FourDefectInteractionExamples}(a) but with $n_2 = 1$.
This changes the predicted velocity of the defects to annihilate in opposite directions, which occurs in the corresponding simulation shown in Supplemental Movie 9.
We can understand the motion from the solitons that appear between the defects in Fig.~\ref{Sfig:FourDefectInteractionExamples}, which are localized distortions that can be removed by defect annihilation.

If the four defects are rearranged so that $x^+ = 0.25$ and $x^- = -0.25$ for both positive and negative defects, the resulting configuration and velocity prediction with $n_1 = n_2 = 0$ is shown in Fig.~\ref{Sfig:FourDefectInteractionExamples}(c).
In this case, the predicted velocity of the defects is that of quick annihilation, borne out in the simulation shown in Supplemental Movie 10, while the Coulomb force that would be experienced by charges in this configuration is zero.
Here, as in the previous two examples, the localized solitons that manifest between defects show why the defects quickly annihilate and leave behind a uniform configuration.
Adding a global soliton, $n_2 = 1$, yields the configuration in Fig.~\ref{Sfig:FourDefectInteractionExamples}(d).
Here it is predicted that the defects do not move, as there is no localized distortion that can be removed from defect annihilation.
Supplemental Movie 11 shows exactly this.

The examples shown in Fig.~\ref{Sfig:FourDefectInteractionExamples} show the important interaction between defect positions and global solitons, and how global solitons can alter the motion of defects with the same positional configuration.
Hence, the history of defect nucleation and annihilation in the system, which can lead to global solitons, is also important to consider when trying to understand the elastic forces felt by the defects.

\section{Supplemental Movies}
\begin{itemize}
    \item Supplemental Movie 1: Simulation of nematic liquid crystal defects in a square, periodic domain of length $L=1$ with initial separation $\Delta z = 0.81$ and $n_1 = n_2 = 0$ corresponding to Fig.~(2)d in the main text.
    The figure on the left shows the initial director configuration.
    The color in the movie is the nematic order parameter $S$ while the white lines show the director $\mathbf{\hat n}$. \href{https://drive.google.com/file/d/17vFeAVVNLJAj1KC5ivFHBURDCjcsC_uf/view?usp=sharing}{Link to video}

    \item Supplemental Movie 2: Simulation of nematic liquid crystal defects in a square, periodic domain of length $L = 1$ with initial separation $\Delta z = 0.2$ and $n_1 = 0$ and $n_2 = 1$ corresponding to Fig.~(3)b in the main text. 
    The figure on the left shows the initial director configuration.
    The color in the movie is the nematic order parameter $S$ while the white lines show the director $\mathbf{\hat n}$. \href{https://drive.google.com/file/d/1d0yoEqOunnHqzbZGvs2vkXzni8IWH0Hi/view?usp=sharing}{Link to video}

    \item Supplemental Movie 3: Simulation of nematic liquid crystal defects in a square, periodic domain of length $L = 1$ with initial separation $\Delta z = 0.1$ and $n_1 = 0$ and $n_2 = 1$ corresponding to Fig.~(3)c in the main text.
    The figure on the left shows the initial director configuration.
    The color in the movie is the nematic order parameter $S$ while the white lines show the director $\mathbf{\hat n}$. \href{https://drive.google.com/file/d/1metLqWb2QVog_vD0jRNajvp0sivxoAh6/view?usp=sharing}{Link to video}
    
    \item Supplemental Movie 4: Simulation of nematic liquid crystal defects in an annulus with radii $r_1 = 1$ and $r_2 = 2$ with initial angular separation $\Delta \phi = 2\pi/5$ and global soliton between defects on the long side of the annulus corresponding to Fig.~(4) in the main text.
    The figure on the left shows the initial director configuration.
    The color in the movie is the nematic order parameter $S$ while the white lines show the director $\mathbf{\hat n}$. \href{https://drive.google.com/file/d/1BIlZyfn_y5JoQZ8Mj-gVvsn9234vgWdX/view?usp=sharing}{Link to video}

    \item Supplemental Movie 5: Simulation of nematic liquid crystal defects in a square, periodic domain of length $L = 1$ with initial separation $\Delta z = 0.9 + 0.9i$ and $n_1 = n_2 = 0$ corresponding to the data in Fig.~\ref{Sfig:DiagonalDefects}(c,d).
    The figure on the left shows the initial director configuration.
    The color in the movie is the nematic order parameter $S$ while the white lines show the director $\mathbf{\hat n}$. \href{https://drive.google.com/file/d/1Jt_zTLhn0sJJefIeXrwz1t39N-tJ5IxQ/view?usp=sharing}{Link to video}

    \item Supplemental Movie 6: Simulation of nematic liquid crystal defects in a square, periodic domain of length $L =1$ with initial separation $\Delta z = 0.8 + 0.4i$ and $n_1 =n_2 = 0$ corresponding to the configuration in Fig.~\ref{Sfig:TwoDefectInteractionExamples}(a).
    The figure on the left shows the initial director configuration.
    The color in the movie is the nematic order parameter $S$ while the white lines show the director $\mathbf{\hat n}$. \href{https://drive.google.com/file/d/15XYLyshZ6GReOpsqIWuuMKAV63nTj63e/view?usp=sharing}{Link to video}

    \item Supplemental Movie 7: Simulation of nematic liquid crystal defects in a square, periodic domain of length $L = 1$ with initial separation $\Delta z = 0.5i$ and $n_1 = 0$ and $n_2 = 1$ corresponding to the configuration in Fig.~\ref{Sfig:TwoDefectInteractionExamples}(b).
    The figure on the left shows the initial director configuration.
    The color in the movie is the nematic order parameter $S$ while the white lines show the director $\mathbf{\hat n}$. \href{https://drive.google.com/file/d/1sFDyGSKqvTgeYQYfWfLFSy4UvtmZelwz/view?usp=sharing}{Link to video}

    \item Supplemental Movie 8: Simulation of four nematic liquid crystal defects in a square, periodic domain of length $L = 1$ with initial positive defect locations $z^+ = 0.25 + 0.25i,\, -0.25-0.25i$, initial negative defect location $z^- = 0.25-0.25i,\,-0.25+0.25i$, and $n_1 = n_2 =0$ corresponding to the configuration in Fig.~\ref{Sfig:FourDefectInteractionExamples}(a).
    The figure on the left shows the initial director configuration.
    The color in the movie is the nematic order parameter $S$ while the white lines show the director $\mathbf{\hat n}$. \href{https://drive.google.com/file/d/1mlDUcNEAJNhaKL5C2ZwW2aAinhED4U_M/view?usp=sharing}{Link to video}

    \item Supplemental Movie 9: Simulation of four nematic liquid crystal defects in a square, periodic domain of length $L = 1$ with initial positive defect locations $z^+ = 0.25 + 0.25i,\, -0.25-0.25i$, initial negative defect location $z^- = 0.25-0.25i,\,-0.25+0.25i$, and $n_1 = 0$ and $n_2 =1$ corresponding to the configuration in Fig.~\ref{Sfig:FourDefectInteractionExamples}(b).
    The figure on the left shows the initial director configuration.
    The color in the movie is the nematic order parameter $S$ while the white lines show the director $\mathbf{\hat n}$. \href{https://drive.google.com/file/d/1Dl-WEgKQm0YhjfZ77qF41Iv6otVktkBe/view?usp=sharing}{Link to video}

    \item Supplemental Movie 10: Simulation of four nematic liquid crystal defects in a square, periodic domain of length $L = 1$ with initial positive defect locations $z^+ = 0.25 + 0.25i,\, 0.25-0.25i$, initial negative defect location $z^- = -0.25+0.25i,\,-0.25-0.25i$, and $n_1 = n_2 = 0$ corresponding to the configuration in Fig.~\ref{Sfig:FourDefectInteractionExamples}(c).
    The figure on the left shows the initial director configuration.
    The color in the movie is the nematic order parameter $S$ while the white lines show the director $\mathbf{\hat n}$. \href{https://drive.google.com/file/d/1c9TTm6JAEOs1fBdwygKRM-4Pii8LHxNd/view?usp=sharing}{Link to video}

    \item Supplemental Movie 11: Simulation of four nematic liquid crystal defects in a square, periodic domain of length $L = 1$ with initial positive defect locations $z^+ = 0.25 + 0.25i,\, 0.25-0.25i$, initial negative defect location $z^- = -0.25+0.25i,\,-0.25-0.25i$, and $n_1 = 0$ and $n_2 =1$ corresponding to the configuration in Fig.~\ref{Sfig:FourDefectInteractionExamples}(d).
    The figure on the left shows the initial director configuration.
    The color in the movie is the nematic order parameter $S$ while the white lines show the director $\mathbf{\hat n}$. \href{https://drive.google.com/file/d/1z4GbzD71qCmYA6bjTSf_lgAK00mR-VKI/view?usp=sharing}{Link to video}
    
\end{itemize}

\bibliography{LC}